\documentclass[prl,10pt,twocolumn]{revtex4-1}
\usepackage{graphicx}
\usepackage{amssymb}      
\usepackage{amsmath}

\begin{document}

\title{Critical network effect induces business oscillations in multi-level marketing systems}
\author{Dranreb Earl Juanico}
\affiliation{Department of Mathematics, School of Science and
Engineering, Ateneo de Manila University, Loyola Heights, Quezon
City 1108, Philippines}
\date{\today}

\begin{abstract}
Business-cycle phenomenon has long been regarded as an empirical
curiosity in macroeconomics. Regarding its causes, recent evidence
suggests that economic oscillations are engendered by fluctuations
in the level of entrepreneurial
activity\cite{beamish2010,koellinger2011}. Opportunities promoting
such activity are known to be embedded in social network
structures\cite{eagle2010,granovetter2005}. However, predominant
understanding of the dynamics of economic oscillations originates
from stylised pendulum models on aggregate macroeconomic
variables\cite{minsky1957,sims1980}, which overlook the role of
social networks to economic activity---echoing the so-called
\textsl{aggregation problem} of reconciling macroeconomics with
microeconomics\cite{colander2008,elsner2007}. Here I demonstrate how
oscillations can arise in a networked economy epitomised by an
industry known as multi-level marketing or MLM\cite{albaum2011}, the
lifeblood of which is the profit-driven interactions among
entrepreneurs. Quarterly data (over a decade) which I gathered from
public MLMs reveal oscillatory time-series of entrepreneurial
activity that display nontrivial scaling and
\textsl{persistence}\cite{morales2012,dimatteo2007}. I found through
a stochastic population-dynamic model, which agrees with the notion
of profit maximisation as the organising principle of capitalist
enterprise, that oscillations exhibiting those characteristics arise
at the brink of a critical balance between entrepreneurial
activation and inactivation brought about by a homophily-driven
\textsl{network effect}. Oscillations develop because of stochastic
tunnelling permitted through the destabilisation by noise of an
evolutionarily stable state. The results fit altogether as evidence
to the Burns-Mitchell conjecture that economic oscillations must be
induced by the workings of an underlying ``network of free
enterprises searching for profit"\cite{burns1954}. I anticipate that
the findings, interpreted under a mesoeconomic
framework\cite{dopfer2012}, could open a viable window for
scrutinising the nature of business oscillations through the lens of
the emerging field of network science. Enquiry along these lines
could shed further light into the network origins of the
business-cycle phenomenon.
\end{abstract}

\maketitle

Known widely in literature as \textsl{network marketing}, MLM
executes through embedded social networks its essential business
functions such as goods distribution, consumption, marketing, and
direct selling\cite{albaum2011,coughlan1998}. That makes MLM a stark
microcosm of a networked economy. One salient yet so far overlooked
feature of MLM dynamics is the aperiodic oscillations in firm size
$N(t)$, quantified by the number of participating entrepreneurs
(Fig.~1). Empirical quarterly firm-size data have been collected
from four public MLMs (Supplementary Data; Supplementary Methods,
S1): NuSkin Enterprises (NUS), Nature Sunshine (NATR), USANA Health
Sciences (USNA), and Mannatech Inc. (MTEX). Publicly-listed firms
are chosen because they are required to disclose accurate business
data on a regular basis. The average revenue (i.e., total revenue
divided by $N$ for any given quarter) does not proportionately rise
with $N$ (Fig.~1c,d), implying that firm-size expansion does not
inevitably translate into revenue growth. Firm size is thus a more
reliable quantifier for entrepreneurial activity than is total
revenue.

The scaling property of the $N(t)$ time-series is examined via Hurst
analysis\cite{morales2012,dimatteo2007} (Methods; Supplementary
Methods, S1). The Hurst exponents, $H(1)$ and $H(2)$, quantify the
scaling of the absolute increments and of the power spectrum,
respectively. If time series were generated by a Wiener process,
such as in the Black-Scholes model\cite{borland2002}, then
$H(1)=0.5$. But $H(1)>0.5$ indicates \textsl{persistence}, i.e.,
changes in one direction usually occur in consecutive periods;
whereas $H(1)<0.5$ suggests \textsl{anti-persistence}, i.e., changes
in opposite directions usually appear in
sequence\cite{dimatteo2007}. Ideally, a single scaling regime means
$H(1)=H(2)$, which applies to time-series generated by unifractal
processes such as the Wiener process and the fractal Brownian
motion. Table~1 presents Hurst exponents for different MLMs.
Generally, $H(1)>0.5$ except for NUS North Asia and NATR with
$H(1)\sim 0.5$ (within standard deviation); and $H(1) = H(2)$ within
standard deviation. Overall, these features of the time-series
suggest that MLM firm dynamics is a non-Wiener unifractal
process\cite{dimatteo2007}. Unifractality implies self-similarity
such that conclusions drawn at one timescale remains statistically
valid at another timescale.

\begin{table}
\centering \caption{Hurst exponents $H(1)$ and $H(2)$ for different
MLMs estimated using the generalised Hurst
method\cite{dimatteo2007,morales2012}. $H(1)>0.5$ indicates
persistence, whereas $H(1)<0.5$ indicates anti-persistence. $H(2)$
values, which are closely related to the scaling of the power
spectrum, are also shown. Generally, $H(2)= H(1)$ within standard
deviation. The standard deviation values are determined from a
pre-testing procedure (Supplementary Methods, S1).}

\begin{tabular}{||c|c|c||}
\hline $MLM$ & $H(1)$ & $H(2)$ \\
\hline \hline
MTEX & $0.8328\pm 0.1140$ & $0.7425\pm 0.1081$ \\
\hline USNA United States & $0.7602\pm 0.1140$ & $0.6823\pm 0.1081$ \\
\hline
USNA Canada & $0.6619\pm 0.1140$ & $0.6013\pm 0.1081$\\
\hline
USNA SEA-Pacific & $0.6383\pm 0.1140$ & $0.6139\pm 0.1081$ \\
\hline
NUS Greater China & $0.6159\pm 0.1098$ & $0.5387\pm 0.1038$ \\
\hline
NATR & $0.5128\pm 0.1032$ & $0.5126\pm 0.0876$ \\
\hline
NUS North Asia & $0.4780\pm 0.1098$ & $0.4789\pm 0.1038$ \\
\hline
\end{tabular}
\label{tab:hursts}
\end{table}

\begin{figure}
\label{fig:empirical} \centering
\includegraphics[totalheight=\columnwidth,angle=-90]{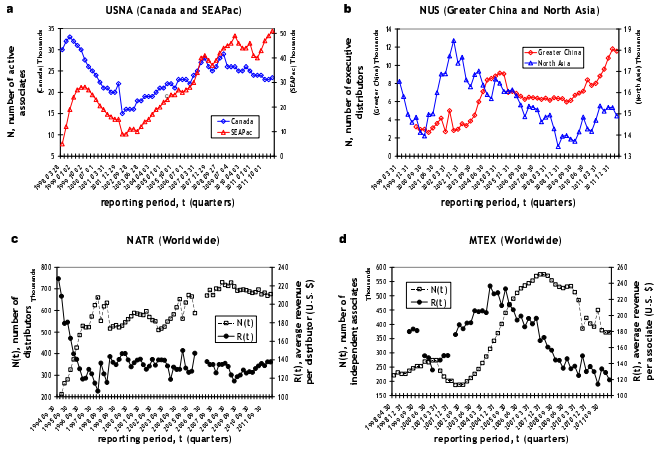}
\caption{Empirical data for different publicly-listed MLM firms.
$N(t)$ is the firm size, $R(t)$ is average revenue per member, and
$t$ is time in quarters. \textbf{a,} USANA Health Sciences (USNA) in
Canada, and Southeast Asia and Pacific (including Australia) from
March 1998 to March 2012. \textbf{b,} NuSkin (NUS) in Greater China
(including Hong Kong) and North Asia (Japan and South Korea) from
March 1999 to March 2012. \textbf{c,} Nature Sunshine (NATR)
worldwide showing $N$ and $R$ from September 1994 to March 2012.
\textbf{d,} Mannatech Inc. (MTEX) worldwide showing $N$ and $R$ from
April 1998 to March 2012. Data sets are provided in Supplementary
Data.}
\end{figure}


An MLM firm is considered as a population of profit-seeking
entrepreneurs. This population exhibits disorder through the
presence of two entrepreneur types distinguished by socio-economic
status (SES). Let $A$ and $B$ denote these types, wherein $A$ has
higher SES than $B$, and $N_A$ and $N_B$ denote their subpopulation
size, respectively. The total population at any given time $t$ is
thus $N(t)=N=N_A+N_B$. Three major processes run the population
dynamics: entrepreneurial activation by recruitment; competitive
inactivation; and catalytic inactivation due to a network effect.
Recruitment is expressed in the following reaction equations:
\begin{eqnarray}\nonumber
A \overset{\mu}\longrightarrow 2A, &\quad B \overset{\mu}\longrightarrow B + A,\\
B \overset{\lambda}\longrightarrow  2B, &\quad A
\overset{\lambda}\longrightarrow A + B, \label{eq:recruitment}
\end{eqnarray}
wherein $\mu$ and $\lambda$ are per-capita rates of recruitment of
types $A$ and $B$, respectively, and $\mu >\lambda$ because $A$'s
higher SES implies a faster rate of entrepreneurial activation
through the support of bigger capital and vaster social resources.
Competitive inactivation occurs due to market overlap, or
\textsl{niche overlap}\cite{gimeno2004}, as participants can go
head-to-head over the same clientele or market. An encounter rate
$\delta$, which can be related to the density of the embedding
social network, quantifies the probability of market overlaps. Thus,
competitive inactivation is:
\begin{equation}
Z + A\overset{\delta}\longrightarrow Z,\quad Z +
B\overset{\delta}\longrightarrow Z\quad
\mathrm{for}\;Z\in\left\{A,B\right\} . \label{eq:competitive}
\end{equation}
Lastly, catalytic inactivation, which denotes the \textsl{network
effect} (Methods), is expressed as:
\begin{equation}
Z
\underset{{\mathbf{MLM}}}{\overset{{\nu\Phi}}{\xrightarrow{\hspace*{1cm}}}}
\varnothing,\quad \mathrm{for}\;Z\in\left\{A,B\right\} .
\label{eq:catalytic}
\end{equation}
The network structure of the MLM catalyses inactivation of existing
participants at the rate $\nu\,\Phi$, where $\Phi
=\dfrac{\sum_{Z\in\{A,B\}}N_Z\left(N_Z-1\right)}{N(N-1)}$ is a
measure of the probability that two members drawn randomly from the
MLM belong to the same type. It has been widely used in literature
as a diversity index\cite{simpson1949}. Due to interconnectedness
and homophily\cite{apicella2012}, the inactivation of one could
(like a contagion) infect another to follow suit.


Combining equations~(\ref{eq:recruitment}--~\ref{eq:catalytic})
results to a Master equation (Supplementary Equation 1) for the
state probability density. Perturbation analysis accounts for the
fluctuations arising from demographic
stochasticity\cite{vankampen2007}. In terms of the system size
$\Omega$ (roughly the size of that part of the overall population
considered \textsl{fit} for entrepreneurial activities) the
following ansatz is made: $N_A=\Omega\alpha+\sqrt{\Omega}\,a$ and
$N_B=\Omega\beta+\sqrt{\Omega}\,b$, where $\alpha$ and $\beta$ are
the average concentrations, and $a$ and $b$ are the magnitude of the
fluctuations of the stochastic variables $N_A$ and $N_B$,
respectively (Supplementary Methods, S2). The highest order in the
expansion expresses the macroscopic rate equations for $\alpha$ and
$\beta$:
\begin{eqnarray}\nonumber
\dot{\alpha}&=\mu(\alpha+\beta)-\nu\,\Phi\,(\alpha+\beta)-\delta\,(\alpha+\beta)\alpha;\\
\dot{\beta}&=\lambda(\alpha+\beta)-\nu\,\Phi\,(\alpha+\beta)-\delta\,(\alpha+\beta)\beta.
\label{eq:macroscopic}
\end{eqnarray}
Meanwhile, the next highest order term gives the Fokker-Planck
equation, or FPE (Supplementary Equation 2), governing the dynamics
of the probability density for the magnitude of the fluctuations.
From the FPE, the expectation values of the stationary fluctuations
are $\left\langle a\right\rangle = \left\langle b\right\rangle=0$,
which supports the interpretation that the deterministic solutions
to~(\ref{eq:macroscopic}) are the correct average values.

The model is nondimensionalised by setting the characteristic
timescale at $t_c=5$~days (Methods). Consequently, the rates can be
squarely related to empirical data by rescaling to appropriate
units. Dimensionless rates take on simplified yet meaningful values:
$\lambda=1-\mu$; $\mu,\nu\in(0,1)$; and
$\Delta=\dfrac{ut_c\delta}{\Omega}\in\left(10^{-4},10^{-3}\right)$.
Bifurcation analysis (Supplementary Methods, S4) of the
nondimensionalised equation~(\ref{eq:macroscopic}) unveils a
bifurcation manifold $\mu=\mu_b(\nu)$, where
\begin{equation}
\mu_b(\nu)=\dfrac{1}{2}+\sqrt{\dfrac{\left(1-\nu\right)^3}{27\nu}},\quad
0<\nu < 1. \label{eq:bifurcation}
\end{equation}
Equation~(\ref{eq:bifurcation}) coincides with an evolutionarily
stable state (ESS) of a population game between the types
(Supplementary Methods, S3--~S4). The fraction of
$A$--~entrepreneurs is $x_A = \overline{x}_A +\Omega^{-1/2}\,\xi$,
where $\overline{x}_A=\dfrac{\alpha}{\alpha+\beta}$, and $\xi$ is
the fluctuation component. Analysis of the second moments from the
FPE shows that the variance diverges as $\left\langle
\xi^2\right\rangle\sim \delta\left|\mu-\mu_b\right|^{-1}$ as
$\mu\rightarrow\mu_b^{-}$ (Supplementary Methods, S5). That is a
signature of criticality through which the ESS, where
$\overline{x}_A = x^*$ and $N = N^* =
\dfrac{1-2\nu\,\Phi(x^*)}{\Delta}$,  is (quite counterintuitively)
destabilised as the bifurcation manifold is approached. This
mechanism is hereby referred as \textsl{stochastic tunnelling}
wherein noise enables the state trajectory to cross a phase barrier
that could not have been otherwise traversed without actively tuning
the bifurcation parameter (Supplementary Methods, S4).

Stochastic tunnelling drives the business oscillations (Fig.~2).
Time series is generated by solving the model using a numerical
technique, known as Gillespie's stochastic simulation
algorithm\cite{gillespie1976}, which directly integrates the master
equation (Supplementary Methods, S6). Diverging variance indeed
allows the solution to wander far enough from the ESS and closer to
an unstable point (UEP) which pushes that solution toward the
boundary state, where $x^*=1$ and $N^*=\dfrac{\mu-\nu\Phi}{\Delta}$
(Fig.~2a). That noise also enables the solution to sling back to the
ESS consequently forming loops in the phase portrait, hence,
oscillations in the time series (Fig.~2b). The time series consist
of upswings associated with increasing diversity and downswings with
decreasing diversity, i.e., $\Phi\rightarrow 1$ as $x_A\rightarrow
1$. The remarkable observation is that recovery from low points of
the series coincide with periods when $A$ is dominant---a case of
the fitter entrepreneurs surviving through
``recessions"\cite{blume2002}.

Profit maximisation is an axiom of capitalist
enterprise\cite{blume2002}. MLM may enhance profitability by
maximising the proportion of $A$~--~entrepreneurs (Supplementary
Methods, S7). Thus, the time-average value $\left\langle
x_A\right\rangle_t$ is examined for various pairs of $\mu$ and $\nu$
which consequently depicts the phase diagram of the model (Fig.~3a).
Business oscillations come about as a result of stochastic
tunnelling through the critical boundary $\mu=\mu_b$. Phase II,
where $A$ stably dominates (Supplementary Fig.~S2b), can be
considered Pareto-optimal as the MLM maximises profitability as a
whole. But high levels of targeted recruitment, i.e.,
$\mu>\dfrac{4}{7}$, are required. Entrepreneurial activation,
however, might in reality be less discriminatory and thus
$\mu\approx\dfrac{1}{2}$, which denotes higher entropy
(Supplementary Discussion). The critical boundary delineates, for
any magnitude $\nu$ of the network effect, the minimum $\mu$ that
promotes long-run dominance of $A$~--~entrepreneurs. Nevertheless, a
stronger network effect tends to frustrate that dominance as
catalytic inactivation increasingly outpaces activation, leading to
degradation of entrepreneurial activity (Supplementary Fig.~S2d).
The ESS at the III-IV boundary (Fig.~3a) is therefore
Pareto-dominated~\cite{webb2007}.


The Hurst maps (Fig.~3b, c) locate where the real MLMs are on the
phase diagram. The Hurst exponents are determined from the same
exact method. Clusters appear in the vicinity of the III-IV
boundary. On these clusters $H(1)$ and $H(2)$ are approximately
between $0.5$ and $0.8$, about the same range of values found in
real MLMs (see Table~1). A correlative plot (Fig.~3d) between $H(1)$
and $H(2)$ further confirms not only agreement between model and
empirical data, but also their unifractality. Overall, these
findings suggest that real-world MLMs are Pareto-dominated economic
systems\cite{elsner2007}, which are operating in an environment
characterised by high entropy, i.e., $\mu\approx\dfrac{1}{2}$, and
by a strong network effect (i.e., $\nu>\mu$).

The study paints an illuminating insight about the nature of MLM
operations. MLMs have been accused in several instances by
discontent participants for ethical violations concerning its
business practices\cite{koehn2001}. The model justifies such
disgruntlement for two reasons. First, that profit is closely
associated with recruitment implies less selective entrepreneurial
activation. Second, that recruitment proceeds through embedded
networks connotes strong network effects. Less-fit entrepreneurs can
join the market in droves but are weeded out too
soon\cite{blume2002} because of the Pareto-dominated nature of the
venture. The feeling of being victimised is thus not at all
surprising.

The mesoeconomic framework (i.e., linking microeconomic foundations
with macroeconomic phenomena\cite{dopfer2012}) puts the present
study in a broader economic context. A more network-dynamic approach
to viewing business cycles is hereby encouraged. Lastly, the
mathematical model could be extended or refined, such as by
generalising the network effect using the H\"{o}lder mean such that
$\Phi(x_A,x_B)=\sqrt[q-1]{x_A^q+x_B^q},\; \forall q>1$
(Supplementary Discussion, Supplementary Fig.~S1); whereas empirical
data of higher temporal resolution may become available in the
future, to further test the implications that came forth.

\begin{figure}
\label{fig:phaseportrait} \centering
\includegraphics[width=\columnwidth]{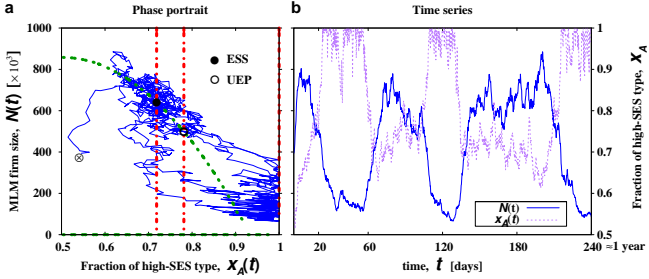}
\caption{Business oscillations in the MLM. The model is simulated
with the following parameters: $\mu=0.57$, $\nu=0.57$,
$\Delta=5\times 10^{-4}$ for a total time of $t=240$~$1-$day periods
$\approx 1$~year. \textbf{a,} Phase portrait showing one stochastic
realization for $N(t)$ (in units $u=10^3$) versus $x_A(t)$;
$\otimes$ marks the initial condition: $N(0)=400$ and $x_A(0)=0.54$.
The dashed curves and lines are the nullclines of the replicator
equations from the evolutionary game (Supplementary Methods, S3),
which intersect at the evolutionarily stable state (ESS, $\bullet$)
and at an unstable equilibrium point (UEP, $\circ$). The trajectory
of the solution forms a loop indicating the oscillations.
\textbf{b,} Time series for $N(t)$ and $x_A(t)$. Hurst analysis
gives $H(1)=0.6668\pm 0.0235$ and $H(2)=0.6564\pm 0.0204$ for the
$N(t)$ series.}
\end{figure}

\begin{figure}
\label{fig:phasediagram}
\includegraphics[totalheight=\columnwidth]{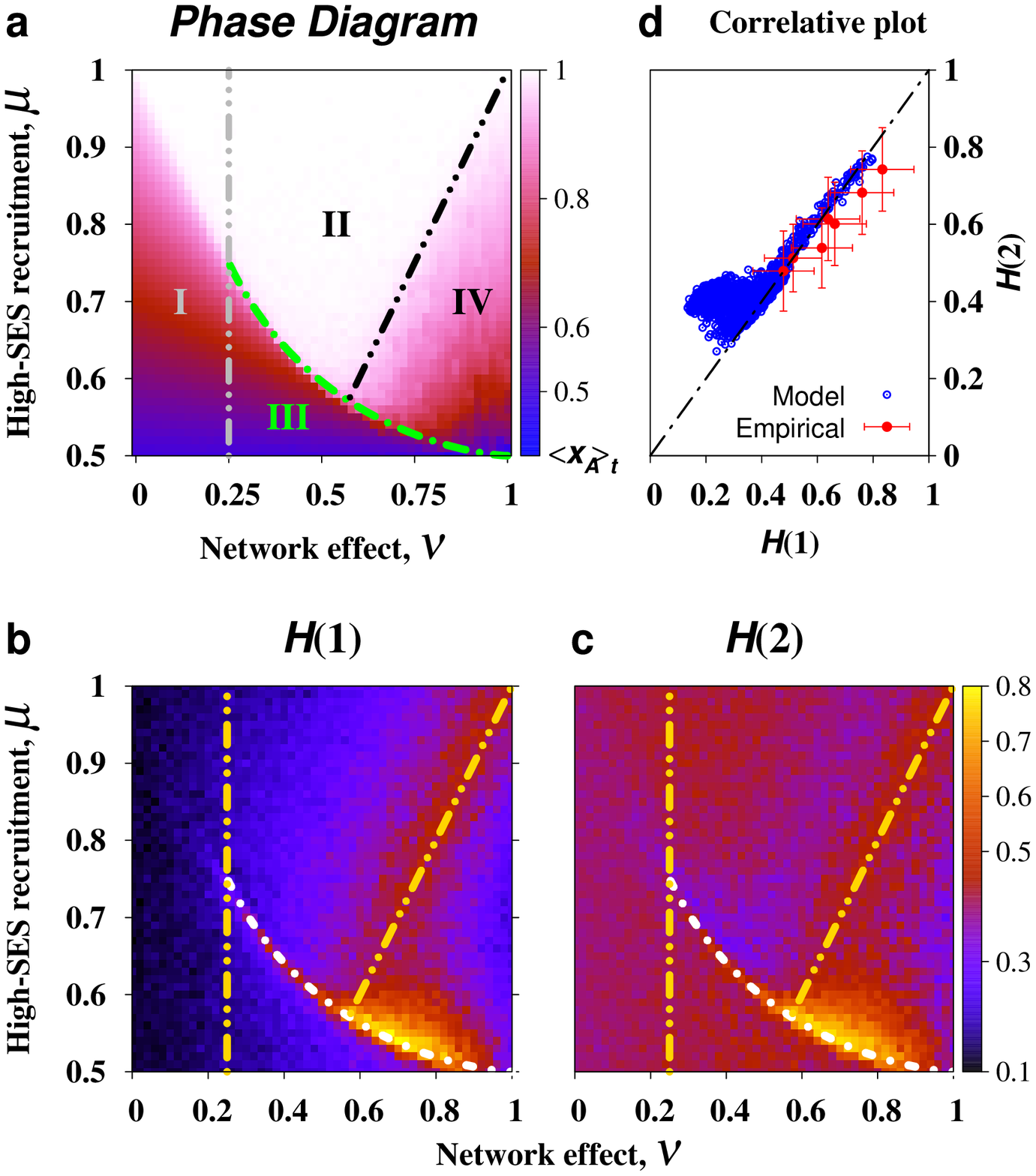}
\caption{Phase diagram and Hurst maps of the MLM. \textbf{a,}
$\left\langle x_A\right\rangle_t$ for different pairs of $\mu$ and
$\nu$ (Resolution: $\Delta\mu\times\Delta\nu = 0.01\times 0.02$).
Phase I represents the regime $0\leq\nu\leq\dfrac{1}{4}$ where the
critical manifold $\mu=\mu_b(\nu)$ is inaccessible due to the
constraint $x_A\leq 1$. Phase II denotes the Pareto-optimal region
$\mu>\mu_b(\nu)$ and $\mu> \nu$ where $x_A\approx 1$. Phase III,
where $\mu <\mu_b$ is similar to phase I except that here the
critical manifold is accessible. Phase IV is where $\mu>\mu_b$ but
$\mu <\nu$ resulting to degradation of $N$ due to $\mu-\nu\,\Phi <
0$ (Supplementary Fig.~2d). \textsl{Phase boundaries}:  I--II,III,
$\nu=\dfrac{1}{4}$; III---II,IV,
 $\mu=\mu_b(\nu)$; and II--IV, $\mu=\nu$. \textbf{b, c,} Map of
the Hurst exponents $H(1)$ and $H(2)$, respectively. The phase
boundaries are superimposed. The data are generated by simulating
the model for $t=420$~$5-$day periods $\approx 8.75$~years with
initial population $N(0)=100$ for different pairs of $\mu$ and $\nu$
(Resolution: $\Delta\mu\times\Delta\nu=0.01\times 0.02$). Each data
pixel is an average of four stochastic realisations. \textbf{d,}
Correlative plot between $H(1)$ and $H(2)$. Empirical data are those
listed in Table~1. The dashed line $H(1)=H(2)$ denotes unifractality
of the time series.}
\end{figure}


\appendix
\section*{Methods}

\subsection{Network effect.}

The \textsl{local network effect}, which is a relatively new idea in
economics\cite{sundararajan2008,galeotti2010,jackson2007,goeree2010},
means that the decision of one entity can influence those by whom
that entity is connected to. Particularly, the network effect
manifests through inactivation as the value of exiting the
enterprise is enhanced through the catalytic action of the
connections between participants.
Homophily\cite{goeree2010,apicella2012} spells that ``a contact
between similar people occurs at a higher rate than among dissimilar
people"\cite{mcpherson2001}, and strongly influences contagions that
diffuse through social links\cite{aral2009}. MLM participants are
thus more likely to connect with others of the same SES,
consequently elevating homogeneity (or depressing diversity) in the
firm. Diversity here is measured by the Simpson index $\Phi$ serving
as a dimensionless potential function minimised when $N_A=N_B$ (at
highest diversity). The network effect is constituted as
$\nu\,\Phi$, for $\nu>0$; hence, the network effect is stronger at
less diversity.

\subsection{Hurst analysis.}
The generalized Hurst method\cite{morales2012,dimatteo2007} has been
coded by one of its authors, T. Aste. The code was downloaded from
Matlab File Exchange website,
\url{http://www.mathworks.com/matlabcentral/fileexchange/30076}, and
was used with default settings in the calculation of $H(1)$ and
$H(2)$.

\subsection{Nondimensionalisation.}
Dimensionless time is defined as $\widehat{t} = t/t_c$, wherein $t_c
= \left(\mu+\lambda\right)^{-1}$. Population numbers are in units of
$u$: $N_A=\widehat{N_A}\,u$ and $N_B=\widehat{N_B}\,u$.
Consequently, equation~(\ref{eq:macroscopic}) becomes:
\begin{widetext}
\begin{eqnarray*}
\dfrac{d\widehat{N_A}}{d\widehat{t}}&=\left(\mu
t_c\right)\widehat{N_A}-\left(\nu
t_c\right)\Phi(\widehat{N_A},\widehat{N_B})\,\left(\widehat{N_A}+\widehat{N_B}\right)-\left(u
t_c\,\delta/\Omega\right)\left(\widehat{N_A}+\widehat{N_B}\right)\widehat{N_A},\\
\dfrac{d\widehat{N_B}}{d\widehat{t}}&=\left(\lambda
t_c\right)\widehat{N_B}-\left(\nu
t_c\right)\Phi(\widehat{N_A},\widehat{N_B})\,\left(\widehat{N_A}+\widehat{N_B}\right)-\left(u
t_c\,\delta/\Omega\right)\left(\widehat{N_A}+\widehat{N_B}\right)\widehat{N_B}.
\end{eqnarray*}
\end{widetext}
Characteristic timescale is chosen at $t_c=5$~days (i.e., $1$~month
$\equiv$ $20$~days). Assuming that the system-size parameter is of
the order, $\Omega\sim 10^6$~individuals, and the unit $u\sim
10^3$~individuals, then setting $10^{-4}<
\Delta=\dfrac{u\,t_c\,\delta}{\Omega} < 10^{-3}$ implies an average
per-capita encounter rate $\delta$ between $1$ and $10$~per month,
which is a reasonable estimate.


\end{document}